\documentclass[aps,pra,twocolumn,floatfix,a4paper,showpacs,superscriptaddress,eqsecnum]{revtex4}
\usepackage{amsmath}
\usepackage{latexsym}
\usepackage{amsfonts}
\usepackage{amssymb}
\usepackage{color}
\usepackage{graphicx}

	\newcommand{\real}{\mathbb R} 
	\newcommand{\half}{\tfrac{1}{2}} 
	\newcommand{\mo}[1]{\left| #1 \right|} 
	\newcommand{\abs}{\mo} 

	\newcommand{\no}[1]{\left\|#1\right\|} 
	\newcommand{\tr}[1]{\textrm{tr}\left[#1\right]} 
	\newcommand{\id}{I} 

	\newcommand{\vg}{\mathbf{g}} 
	\newcommand{\vp}{\mathbf{p}} 
	\newcommand{\vq}{\mathbf{q}} 
	\renewcommand{\vr}{\mathbf{r}} 
	\newcommand{\vx}{\mathbf{x}} 
	\newcommand{\vy}{\mathbf{y}} 
	\newcommand{\vz}{\mathbf{z}} 
	\newcommand{\vsigma}{\boldsymbol{\sigma}} 
	\newcommand{\vnull}{\mathbf{0}}

	\newcommand{\A}{\mathsf{A}}
	\newcommand{\B}{\mathsf{B}}
	\newcommand{\C}{\mathsf{C}}
	\newcommand{\E}{\mathsf{E}}
	\newcommand{\F}{\mathsf{F}}
	\newcommand{\G}{\mathsf{G}}
	\newcommand{\X}{\mathsf{X}}
	\newcommand{\Y}{\mathsf{Y}}
	\newcommand{\Z}{\mathsf{Z}}

	\newcommand{\Lu}{\mathcal{L}} 

	\newcommand{\GI}{\mathcal{G}}
	\newcommand{\EI}{\mathcal{E}}
	\newcommand{\FI}{\mathcal{F}}
	\newcommand{\AI}{\mathcal{A}}
	\newcommand{\BI}{\mathcal{B}}
	\newcommand{\CI}{\mathcal{C}}
	\newcommand{\XI}{\mathcal{X}}
	\newcommand{\YI}{\mathcal{Y}}
	\newcommand{\ZI}{\mathcal{Z}}
	\newcommand{\dist}{\mathfrak{D}} 
	\newcommand{\sdist}{\mathfrak{W}} 

\begin{document}

\title[Approximating incompatible von Neumann measurements]{Approximating incompatible von Neumann measurements simultaneously}

\author{Teiko Heinosaari}
\affiliation{Niels Bohr Institute, Blegdamsvej 17, 2100 Copenhagen, Denmark}
\email{heinosaari@nbi.dk}

\author{Maria Anastasia Jivulescu}
\affiliation{Niels Bohr Institute, Blegdamsvej 17, 2100 Copenhagen, Denmark}
\affiliation{Department of Mathematics,  University Politehnica Timisoara, P-ta Victoriei Nr.~2, 300006 Timisoara, Romania}
\email{jivulescu@nbi.dk}

\author{Daniel Reitzner}
\affiliation{Institute of Physics, Slovak Academy of Sciences, D\'ubravsk\'a cesta 9, 845 11 Bratislava, Slovakia}
\email{daniel.reitzner@savba.sk}

\author{Mario Ziman}
\affiliation{Institute of Physics, Slovak Academy of Sciences, D\'ubravsk\'a cesta 9, 845 11 Bratislava, Slovakia}
\affiliation{Faculty of Informatics, Masaryk University, Botanick\'a 68a, Brno, Czech Republic}
\email{ziman@savba.sk}

\begin{abstract}
We study the problem of performing orthogonal qubit measurements simultaneously. 
Since these measurements are incompatible, one has to accept additional imprecision.
An optimal joint measurement is the one with the least possible imprecision.
All earlier considerations of this problem have concerned only joint measurability of observables, while in this work we also take into account conditional state transformations (i.e.,~instruments).
We characterize the optimal joint instrument for two orthogonal von Neumann instruments as being the L\"{u}ders instrument of the optimal joint observable. 
\end{abstract}


\maketitle

\section{Introduction}\label{sec:intro}

It is a fundamental fact of quantum theory that there exist pairs of incompatible measurements.
The simplest example is a pair of (ideal) spin component measurements in different directions.
These measurements cannot be measured jointly using a single device.

The existence of incompatible measurements (i.e., impossibility of certain joint measurements) is linked with some other impossible tasks, such as cloning and teleportation.
For each impossible device, one can study its best approximative substitute.
This kind of optimal \emph{possible} device then gives an absolute bound for the error one has to face in any attempt to build 
the impossible device. Evidently, this kind of quantitative bound on the error can tell us much more than just a plain statement 
of impossibility.

The question of approximate joint measurements of two sharp qubit observables (e.g.,~spin-1/2 components) was first studied in \cite{Busch86}. In recent years this topic has been investigated from several different aspects.  The Mach-Zehnder interferometric setup  was analyzed in \cite{BuSh06,LiLiYuCh09} from the point of view of joint measurements. 
Various trade-off relations concerning joint approximations were derived in \cite{KuSaUe07,SaUe08,BuHe08,BrAnBa09}. Characterizations of all jointly measurable two-outcome qubit observables were determined in \cite{StReHe08,YuLiLiOh08,BuSc10}. 
A connection between the CHSH Bell inequality  \cite{ClHoShHo69} and the bound on joint qubit measurements was observed in \cite{AnBaAs05}, and in \cite{WoPeFe09} it was shown that every pair of two-outcome observables being not jointly measurable enables the violation of the CHSH Bell inequality.
The relationship between cloning of observables and joint measurements was investigated in \cite{FePa07}.

In the current work we study the question of approximate joint measurement of two sharp qubit measurements from a different perspective.
In earlier works, discussion has concerned only joint measurability of \emph{observables}.
In this work we extend the problem to a joint measurability of \emph{instruments}. 
In other words, we consider approximations not only to measurement outcome probabilities but also to conditional state transformations.
One of our main results is the characterization of the optimal joint instrument for two orthogonal von Neumann instruments.

This paper is organized as follows.
In Sec.~\ref{sec:coexistence} we explain the two different levels of compatibility.
Some useful details on joint observables are presented in Sec.~\ref{sec:joint-observable}.
In Sec.~\ref{sec:joint-instrument} a general form for joint instruments is derived.
The optimal approximate joint instrument for two von Neumann instruments is then characterized in Sec.~\ref{sec:optimal}.
Finally, in Sec.~\ref{sec:three} we discuss the case of three von Neumann measurements.

\section{Two levels of incompatibility}\label{sec:coexistence}

Compatibility of quantum measurements has different meanings depending on what we take into consideration.
In particular, two measurements can be compatible if we care only about the bare measurement outcome statistics, but fail to be compatible if we take into account the dynamics of the measurements.
This fact is the motivation for the current investigation and in the following we explain this twofold meaning in detail.

\begin{figure}
\begin{center}
a) \includegraphics[scale=0.75]{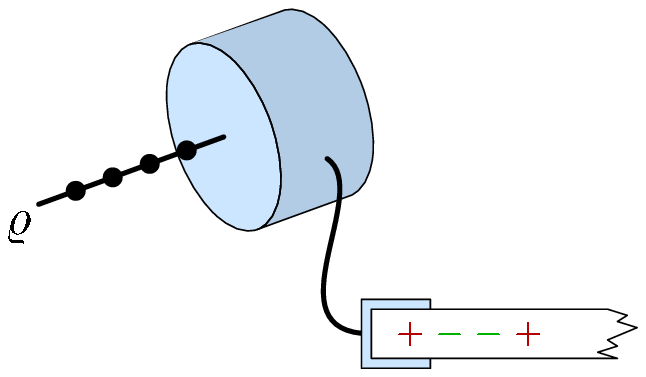}\vskip5mm b) \includegraphics[scale=0.75]{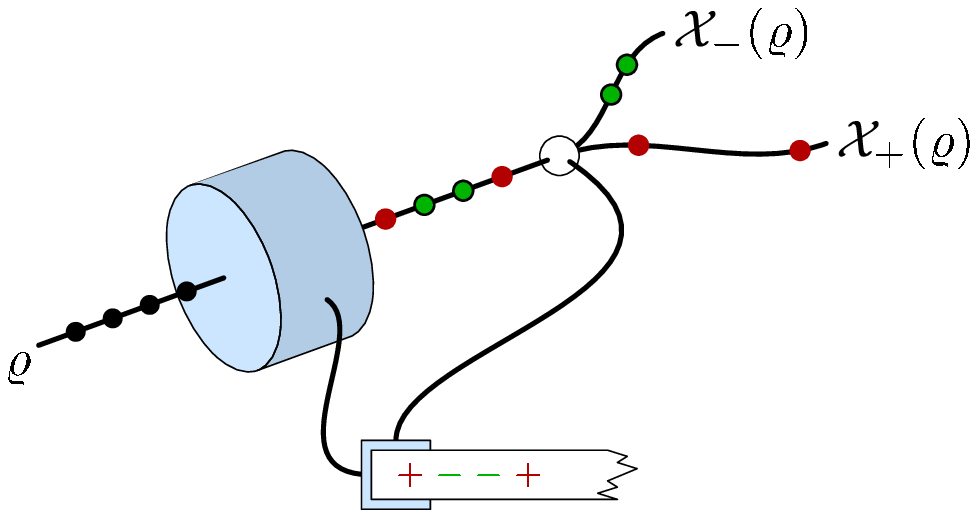}
\end{center}
\caption{\label{fig:x}(a) An observable describes the measurement outcome statistics, whereas (b) an instrument describes both the measurement outcome statistics and the conditional output states with possible separation in accordance with the classical outcomes.}
\end{figure}

In particular, let us consider two sharp observables $\X$ and $\Y$ on a qubit system.
These can be, for instance, spin component measurements on a spin-$\half$ system.
The observables $\X$ and $\Y$ are described by the selfadjoint operators $\sigma_\vx = \vx\cdot\vsigma$ and $\sigma_\vy=\vy\cdot\vsigma$, where $\vsigma=(\sigma_1,\sigma_2,\sigma_3)$ are the Pauli matrices and $\vx,\vy$ are unit vectors.
Alternatively, and for our purposes more conveniently, these observables can be described by \emph{projection valued measures} (PVMs).
Then $\X$ and $\Y$ are identified as mappings from a set of measurement outcomes to projectors
\begin{equation*}
\pm 1 \mapsto \X(\pm 1)=\half (\id \pm \sigma_\vx) \, , \quad \pm 1 \mapsto \Y(\pm 1)=\half (\id \pm \sigma_\vy) \, .
\end{equation*}
A measurement of $\X$ (similarly $\Y$) gives either a result up (+1) or down (-1); see Fig.~\ref{fig:x}a.
For instance, if the system is in a state $\varrho$, then the probability of getting the outcome $1$ in a measurement of $\X$ is $\tr{\varrho \X(1)}$.
The operator $\sigma_\vx$ gives the average value of the $\X$ measurement, which means that the formula,
\begin{equation*}
\tr{\varrho\X(1)}-\tr{\varrho\X(-1)}=\tr{\varrho\sigma_\vx},
\end{equation*}
holds for all states $\varrho$.

In our following investigation we assume that the unit vectors $\vx$ and $\vy$ are orthogonal.
This is equivalent to the condition that $\tr{\X(i)\Y(j)}=\half$ for every $i,j=\pm 1$. 
Hence, certain predictability of one outcome of $\X$ implies that both outcomes of $\Y$ are equally likely, and vice versa.
This relation is usually referred to as \emph{(value) complementarity} \cite{BuLa95}.

Our assumption on the orthogonality of $\vx$ and $\vy$ means, in particular, that the observables $\X$ and $\Y$ do not commute, i.e., $\X(i)\Y(j)\neq \Y(j)\X(i)$.
The noncommutativity implies the impossibility of performing their joint measurement.
Therefore, we need to choose whether we measure $\X$ or $\Y$, their simultaneous measurement being impossible.

It is possible to approximate $\X$ and $\Y$ with a pair of jointly measurable observables $\A$ and $\B$ described by \emph{positive operator valued measures} (POVMs) \cite{PSAQT82,OQP97}.
An essential fact is that for POVMs (unlike for PVMs) commutativity is not a necessary condition for joint measurability.

Suppose we want to approximate $\X$ and $\Y$ equally well.
Then a class of approximating observables, parametrized by a number $0\leq\eta\leq 1$, is defined by
\begin{equation*}
\A(\pm 1) = \frac{1}{2} \left( I \pm \eta \sigma_\vx\right)  \, , \quad \B(\pm 1) = \frac{1}{2} \left( I \pm \eta \sigma_\vy \right) \, .
\end{equation*}
The number $\eta$ quantifies how close $\A$ and $\B$ are to $\X$ and $\Y$, respectively.

In the limiting case $\eta=1$ we have $\A=\X$ and $\B=\Y$, however, in such case $\A$ and $\B$ are not jointly measurable.
It was shown in \cite{Busch86} that $\A$ and $\B$ have a joint measurement if and only if $\eta\leq\frac{1}{\sqrt{2}}$.
Therefore, we fix $\eta=\frac{1}{\sqrt{2}}$ and $\A,\B$ are then the optimal jointly measurable approximations to $\X,\Y$.

At this point one may wonder whether the $\eta$-parametrized class of observables leads to the best approximation, or perhaps some modification gives a better approximation (while preserving joint measurability).
However, in \cite{KuSaUe07,BuHe08} it has been proved that any modification to $\A$ and $\B$ leads either to a worse approximation or lack of joint measurability.

A joint observable for the observables $\A$ and $\B$ is defined as a POVM $\G$ with four outcomes corresponding to four possible pairs of $\A$ and $\B$ outcomes, $(\pm 1,\pm 1)$.
It is required that the measurement outcome statistics for $\A$ ($\B$) measured alone can be obtained from the joint observable by disregarding (summing through all possible) outcomes for $\B$ ($\A$).
Hence, the defining condition for $\G$ is that
\begin{subequations}
\label{eq:marginals}
\begin{eqnarray}
\A(\pm 1) &=& \G(\pm 1,1) + \G(\pm 1,-1) \, ,  \label{eq:A-marginal} \\
\B(\pm 1) &=& \G(1,\pm 1) + \G(-1,\pm 1) \, . \label{eq:B-marginal}
\end{eqnarray}
\end{subequations}
In other words, $\A$ and $\B$ are \emph{marginals} of $\G$.
A possible choice is
\begin{subequations}
\label{eq:G}
\begin{eqnarray}
\G(+1,\pm 1) &=& \frac{1}{4} \left[\id+\frac{1}{\sqrt{2}} (\sigma_\vx \pm \sigma_\vy)\right] \, , \label{eq:G1} \\
\G(-1,\pm 1) &=& \frac{1}{4} \left[\id  - \frac{1}{\sqrt{2}} (\sigma_\vx \mp \sigma_\vy)\right] \, . \label{eq:G2}
\end{eqnarray}
\end{subequations}
It is easy to verify that $\G$ indeed fulfills the requirements \eqref{eq:marginals} and that each $\G(\pm 1,\pm 1)$ is a positive operator.
Various ways to realize $\G$ and other related measurements have been discussed (e.g.,~in \cite{BuSh06,Busch87}).

So far, our discussion has concerned only joint measurability of observables (i.e., compatibility of measurement outcome probabilities). There is also another level of compatibility, arising from the fact that a (nontrivial) quantum measurement necessarily affects the state of the measured system. Thus, each measurement outcome has an associated operation, which is mathematically described as a completely positive trace-nonincreasing mapping on the set of states. The collection of all these operations forms an \emph{instrument} \cite{QTOS76}.

The standard measurement for a discrete sharp observable is the so-called \emph{von Neumann measurement}. The corresponding instrument, which we call the von Neumann instrument, has a very simple form. In our case, the von Neumann instruments $\XI$ and $\YI$ associated with the sharp observables $\X$ and $\Y$, respectively, are given by
\begin{subequations}
\label{eq:vn}
\begin{eqnarray}
\XI_{\pm}(\varrho) &=& \X(\pm 1) \varrho \X(\pm 1) \, ,\\
\YI_{\pm}(\varrho) &=& \Y(\pm 1) \varrho \Y(\pm 1) \, .
\end{eqnarray}
\end{subequations}
For instance, if the system is in a state $\varrho$ and a measurement of $\X$ gives the outcome $1$, then the unnormalized output state is $\XI_+(\varrho)$; see Fig.~\ref{fig:x}(b).
We can also write
\begin{equation*}
\XI_{+}(\varrho) = \X(1) \varrho \X(1) = \tr{\varrho \X(1)} \X(1) \, ,
\end{equation*}
which shows that the normalized output state is $\X(1)$. 

Since $\X$ and $\Y$ are not jointly measurable, none of their instruments can be jointly measurable. 
In particular, there is no measurement scheme which would realize both $\XI$ and $\YI$. 
Therefore, if we want to realize the instruments $\XI$ and $\YI$ in a single measurement scheme, we need to approximate them.

In the case of the approximating observables $\A,\B$ the von Neumann instruments are commonly replaced by \emph{L\"uders instruments} $\Lu^\A$ and $\Lu^\B$, defined as
\begin{subequations}
\label{eq:luders}
\begin{eqnarray}
\Lu^\A_{\pm}(\varrho) &=& \sqrt{\A(\pm 1)} \varrho \sqrt{\A(\pm 1)} \, ,\\
\Lu^\B_{\pm}(\varrho) &=& \sqrt{\B(\pm 1)} \varrho \sqrt{\B(\pm 1)} \, .
\end{eqnarray}
\end{subequations}
For the operator $\A(1)$, the square root $\sqrt{\A(1)}$ takes the form,
\begin{equation*}
\sqrt{\A(1)} = \sqrt{\frac{1+\eta}{2}} \ \X(1) +  \sqrt{\frac{1-\eta}{2}} \ \X(-1) \, .
\end{equation*}
Hence, we see that
\begin{equation*}
\Lu^\A_+(\varrho) = \half (1+\eta)\XI_+(\varrho) + \textrm{other terms} \, .
\end{equation*}
In this way, we can understand the $\Lu^\A$ measurement as an approximate version of the $\XI$ measurement. 
We refer to \cite{Com} for a convenient summary of the L\"uders instrument in general.

In the limiting case $\eta=1$ when $\A=\X$ and $\B=\Y$ the formulas \eqref{eq:vn} and \eqref{eq:luders} coincide. Hence, we would expect that the L\"uders instruments of $\A$ and $\B$ are good approximations to the von Neumann instruments of $\X$ and $\Y$. 
Here, however, we face a problem. It was shown in \cite{HeReStZi09} that two L\"uders operations $\Lu^\A_+$ and $\Lu^\B_+$ are jointly measurable if and only if \footnote{Here we make use of the usual notation that $\A\geq\B$ stands for the operator $\A-\B$ being positive.} either $\A(1)+\B(1)\leq I$ or $\B(1)=b \A(1)$ for some $0\leq b \leq 1$.
Since neither of these two conditions holds in our situation, we find that the L\"uders instruments $\Lu^\A$ and $\Lu^\B$  cannot be realized in a single experimental setup. Therefore, they do not provide the jointly measurable approximations that we are looking for.

We conclude that the obvious replacements for the von Neumann instruments of $\X$ and $\Y$, namely the L\"uders instruments of $\A$ and $\B$, are not jointly measurable, although the observables $\A$ and $\B$ are.
On the other hand, joint measurability of $\A$ and $\B$ implies that they have \emph{some} jointly measurable instruments.
In fact, every instrument implementing a joint observable of $\A$ and $\B$ gives instruments for $\A$ and $\B$ as its marginals.
In the following we will characterize the joint instrument which gives the best approximations for the von Neumann instruments of $\X$ and $\Y$.

\section{Joint observable}\label{sec:joint-observable}

In this section we derive some useful properties of the joint observable $\G$, defined in \eqref{eq:G}.
Let us first make a general observation.
Suppose that $\A$ and $\B$ would have a second joint observable $\G'$.
Then, also all the convex combinations $\lambda \G + (1-\lambda) \G'$, $0<\lambda< 1$, defined as ($i,j=\pm 1$)
\begin{equation*}
[\lambda \G + (1-\lambda) \G'](i,j) = \lambda \G(i,j) + (1-\lambda) \G'(i,j) \, ,
\end{equation*}
are joint observables of $\A$ and $\B$.
This leads to the conclusion that $\A$ and $\B$ either have a unique joint observable or uncountably many different joint observables. The case under investigation falls, luckily, into the first class. This is essential for our investigation as it crucially limits the search for optimal joint instruments.

\begin{figure}
\begin{center}
\includegraphics[scale=0.75]{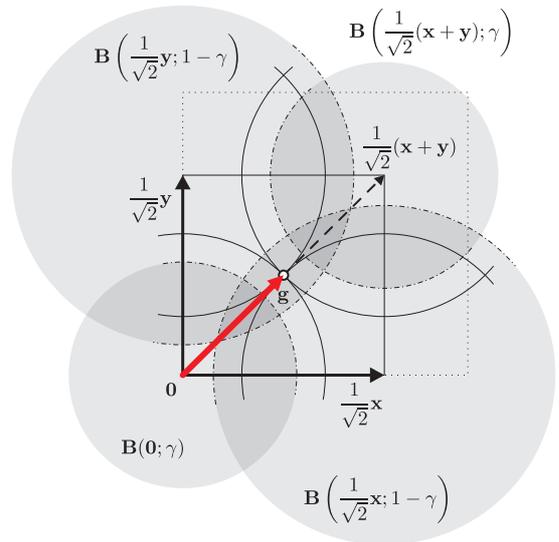}
\end{center}
\caption{\label{fig:2-uniqueness}Visual representation of the four-ball intersection determining the existence of joint observable for $\A$ and $\B$ parametrized by $\gamma$ and vector $\vg$. If $\gamma=1/2$, all four balls intersect in a single point (white circle) defining the joint observable (solid arcs). If $\gamma$ differs from $1/2$ (e.g.~$\gamma=0.4$ as in the figure), two opposite circles intersect, while the other two do not; in such a case there is no joint observable with given $\gamma$ (dot-dashed arcs).}
\end{figure}

To see that $\G$ is a unique joint observable for $\A$ and $\B$, we first notice that any joint observable $\G'$ for $\A$ and $\B$ is completely determined by a single operator, say $\G'(1,1)$.
The other operators are then recovered from the marginal conditions \eqref{eq:marginals}.
Since $\G'(1,1)$ is a positive operator, we can write it as
\begin{equation*}
\G'(1,1)=\half (\gamma\id + \vg\cdot\vsigma) \, , 
\end{equation*}
where $\gamma\geq 0$ and $\vg\in\real^3$.
As noticed in \cite{Busch86}, the conditions for $\G'(1,1)$ to define a joint observable for $\A$ and $\B$ are the following operator inequalities:
\begin{eqnarray*}
O \leq \G'(1,1) &\leq& \A(1) \, ,\\
\A(1)+\B(1)-\id \leq \G'(1,1) &\leq& \B(1) \, .  
\end{eqnarray*}
These are equivalent to the requirement that the vector $\vg$ is in the intersection of four balls:
\begin{multline*}
\vg\in \mathbf{B}(\vnull;\gamma) \cap \mathbf{B}\left(\frac{1}{\sqrt{2}}\vx;1-\gamma\right)\\
\cap \mathbf{B}\left(\frac{1}{\sqrt{2}}\vy;1-\gamma\right) \cap \mathbf{B}\left(\frac{1}{\sqrt{2}}(\vx+\vy);\gamma\right) \, .
\end{multline*}
Since $\vx$ and $\vy$ are orthogonal unit vectors, the intersection is nonempty only if $\gamma=\half$, and in that case $\vg=\frac{1}{2\sqrt{2}}(\vx+\vy)$; see Fig.~\ref{fig:2-uniqueness}.
This means that $\G'=\G$, proving that there is only one joint observable for $\A$ and $\B$.

It may be worth emphasizing that the joint observable of two observables is unique only in special cases.
For instance, if instead of taking $\eta=\frac{1}{\sqrt{2}}$ we would have chosen a smaller number in the definition of $\A$ and $\B$, then they would have infinitely many joint observables.
This is as well evident from Fig.~\ref{fig:2-uniqueness}.

Let us denote by $\mu_{\pm\pm}(\varrho)\equiv\tr{\varrho\G(\pm 1,\pm 1)}$ the probabilities observed in
the $\G$ measurement. If a state $\varrho$ is written as $\varrho=\half (\id + \vr\cdot\vsigma)$, then
\begin{equation*}
\mu_{jk}(\varrho) = \frac{1}{4} \left[ 1 + \frac{1}{\sqrt{2}} (j\vr\cdot\vx + k\vr\cdot\vy) \right] \, , \quad j,k=\pm \, .
\end{equation*}
It is now straightforward to see that
\begin{equation*}
\mu_{+\mp}(\varrho) + \mu_{-\pm}(\varrho)=\frac{1}{2}
\end{equation*}
for all states $\varrho$.
This implies that the whole probability distribution is actually determined only by two numbers [e.g.,~$\mu_{++}(\varrho)$ and $\mu_{+-}(\varrho)$].
We further notice that the numbers $\mu_{++}(\varrho)$ and $\mu_{+-}(\varrho)$ satisfy 
\begin{equation*}
\label{eq:convex2D}
\left(\mu_{++}(\varrho)-\frac{1}{4}\right)^2+\left(\mu_{+-}(\varrho)-\frac{1}{4}\right)^2 \leq \left(\frac{\no{\vr}}{4}\right)^2 \, .
\end{equation*}
This inequality characterizes the convex set of all possible probability distributions in the range of $\G$.

\section{Joint instrument}\label{sec:joint-instrument}

An instrument $\GI$ implementing the joint observable $\G$ consists of four operations  $\GI_{jk}$, $j,k=\pm$, satisfying
\begin{equation*}
\tr{\GI_{jk}(\varrho)}=\tr{\varrho \G(j,k)}\,.
\end{equation*}
Due to the simple structure of $\G$, we can characterize all its instruments $\GI$ in an uncomplicated way. Let us, for a moment, concentrate on $\GI_{++}$, an operation associated with the outcome combination $(1,1)$.

We first observe that $2\G(1,1)=:R$ is a one-di\-men\-sional projection. Let $\{K_\ell\}$ be the set of Kraus operators for $\GI_{++}$, so that
\begin{eqnarray*}
\GI_{++}(\varrho) &=& \sum_\ell K_\ell \varrho K^\ast_\ell \, ,\\
\sum_\ell K_\ell^\ast K_\ell &=& \G(1,1)=\half R \, .
\end{eqnarray*}
The last equation implies that for each $\ell$, we have $2K_\ell^\ast K_\ell \leq R$.
Since $R$ is a one-dimensional projection, there is a number $0< k_\ell \leq 1$ such that $2K_\ell^\ast K_\ell = k_\ell R$.
Clearly, $\sum_\ell k_\ell =1$.
Let $K_\ell = U_\ell \abs{K_\ell}$ be the polar decomposition of $K_\ell$.
Here $U_\ell$ is a unitary operator and
\begin{equation*}
\abs{K_\ell} = \sqrt{K_\ell^\ast K_\ell} =  \sqrt{k_\ell/2} R \, .
\end{equation*}
For every state $\varrho$, we then get
\begin{eqnarray*}
K_\ell \varrho K_\ell^\ast &=& \half k_\ell  U_\ell R \varrho R U_\ell^\ast = \half k_\ell \tr{\varrho R} U_\ell R U_\ell^\ast\\
&=& \tr{\varrho \G(1,1)} k_\ell U_\ell R U_\ell^\ast
\end{eqnarray*}
and hence
\begin{equation*}
\GI_{++}(\varrho) = \tr{\varrho \G(1,1)} \sum_\ell k_\ell U_\ell R U_\ell^\ast \, .
\end{equation*}
Each $U_\ell R U_\ell^\ast$ is a one-dimensional projection and the convex sum, 
\begin{equation*}
\sum_\ell k_\ell U_\ell R U_\ell^\ast  \equiv \xi_{++},
\end{equation*}
is therefore a state.

A similar calculation can be performed for the other three operations separately.
Hence, we conclude that an instrument $\GI$ implementing  $\G$ is determined by four states $\xi_{\pm\pm}$, and the
corresponding operations are given by
\begin{equation}\label{eq:atomic_instrument}
\GI_{\pm\pm}(\varrho) = \tr{\varrho \G(\pm 1, \pm 1)} \xi_{\pm\pm} = \mu_{\pm\pm}(\varrho) \xi_{\pm\pm}\, .
\end{equation}
As we have seen, this simple structure of the instruments implementing $\G$ is due to the fact that each element $\G(i,j)$ is a rank-1 operator.

Finally, let us emphasize that the probabilities $\mu_{\pm\pm}(\varrho)$ are fixed since $\GI$ implements the observable $\G$.
The freedom we have is only in the choice of the four states $\xi_{\pm\pm}$.

\section{Optimal approximation}\label{sec:optimal}

\subsection{Distance between operations}\label{sec:distance}

We are seeking for the best simultaneous approximation to the von Neumann instruments associated with $\X$ and $\Y$.
We therefore perform a measurement of $\G$, which is described by an instrument $\GI$ of the form \eqref{eq:atomic_instrument}.

In a similar way as $\G$ gives $\A$ and $\B$ as its marginals, $\GI$ determines marginal instruments $\AI$ and $\BI$.
Our aim is that the following approximations should be as close as possible:
\begin{eqnarray*}
\label{eq:approx_P+}\AI_{+} \equiv \GI_{++} + \GI_{+-} & \sim & \XI_{+}\,,\\
\label{eq:approx_Q+}\BI_{+} \equiv \GI_{++} + \GI_{-+} & \sim & \YI_{+}\,,\\
\label{eq:approx_P-}\AI_{-} \equiv \GI_{--} + \GI_{-+} & \sim & \XI_{-}\,,\\
\label{eq:approx_Q-}\BI_{-} \equiv \GI_{--} + \GI_{+-} & \sim & \YI_{-}\,.
\end{eqnarray*}

As we are already using the unique joint observable $\G$ of the optimal approximating observables $\A$ and $\B$, the measurement outcome probabilities are set and do not depend on the choice of $\GI$. Therefore, in order to quantify the distance between a given approximation and the corresponding von Neumann instrument, it is enough to compare the normalized output states.

There are various options for how to quantify the distance between Hilbert space operators. However, when considering the distance between density operators it is natural to choose the one induced by the trace norm. Operationally, it quantifies the optimal probability with which the states can be discriminated in a single run of the experiment (i.e., by observing a single experimental click \cite{QDET76}). 

The distance exhibiting the difference between the output states for
a given pair of operations can be utilized to induce a distance between 
the instruments. In particular, in what follows we will analyze
the average distance over all input states (Sec.~\ref{sec:avgoptimal}) and the worst-case distance (Sec.~\ref{sec:supoptimal}).
Our interest is to minimize their values for all outcomes.

If we measure $\X$ and obtain the outcome $1$, then the output state is $\X(1)$.
On the other hand, if we measure $\A$ and obtain the outcome $1$, then the output state is 
\begin{eqnarray*}
\frac{\AI_{+}(\varrho)}{\tr{\AI_+(\varrho)}} &=& \frac{\mu_{++}(\varrho)}{\mu_{++}(\varrho)+\mu_{+-}(\varrho)}\xi_{++}\\
&&+\frac{\mu_{+-}(\varrho)}{\mu_{++}(\varrho)+\mu_{+-}(\varrho)}\xi_{+-} \, .
\end{eqnarray*}
The trace distance of the approximation $\AI_{+}$ from the desired operation $\XI_{+}$, given that the input state is $\varrho$, is thus
\begin{eqnarray*}
d^{\A,\X}_{+}(\varrho) :=  \no{\frac{\AI_{+}(\varrho)}{\tr{\AI_+(\varrho)}} - \X(1)} \, ,
\end{eqnarray*}
where the norm on the right-hand side is the trace norm.

\subsection{Optimal approximations under average distance}\label{sec:avgoptimal}

Assigning Bloch vectors $\vq_{\pm\pm}\in\real^3$ to the states $\xi_{\pm\pm}$, the distance $d^{\A,\X}_{+}(\varrho)$ can be written in the form,
\begin{equation}
d^{\A,\X}_{+}(\varrho) = \no{\frac{\vq_{++}+\vq_{+-}}{2} + f(\varrho) \frac{\vq_{++}-\vq_{+-}}{2}-\vx} \, ,\label{eq:d-bloch}
\end{equation}
where $\vr=(r_x,r_y,r_z)$ is the Bloch vector corresponding to $\varrho$, $f(\varrho)=f(r_x,r_y)=r_y/(\sqrt{2}+r_x)$ and the norm is the Euclidean norm in $\real^3$. 
Similarly we get
\begin{eqnarray*}
d^{\B,\Y}_{+}(\varrho) &=& \no{\frac{\vq_{++}+\vq_{-+}}{2} + f(\varrho) \frac{\vq_{++}-\vq_{-+}}{2}-\vy} \\
d^{\A,\X}_{-}(\varrho) &=& \no{\frac{\vq_{--}+\vq_{-+}}{2} + f(\varrho) \frac{\vq_{--}-\vq_{-+}}{2}+\vx} \\
d^{\B,\Y}_{-}(\varrho) &=& \no{\frac{\vq_{--}+\vq_{+-}}{2} + f(\varrho) \frac{\vq_{--}-\vq_{+-}}{2}+\vy}
\end{eqnarray*}

Since both the vectors $\vx,\vy$ have a vanishing $z$ com\-po\-nent, 
it follows that setting the $z$ component of our choice of
Bloch vectors $\vq_{\pm\pm}$ to zero decreases the distances. 
Hence, in optimization tasks we can restrict ourselves to vectors $\vq_{\pm\pm}$ 
with the vanishing $z$ component.

We denote by $\left\langle\ \cdot\ \right\rangle_{\varrho}$ the normalized integration over the Bloch ball ${\mathbb B}=\{\vr\in{\mathbb R}^3:\no{\vr}^2=r_x^2+r_y^2+r_z^2\leq 1\}$ representing the state space of a qubit.
Hence, for a function $F$ defined on the state space we have
\begin{equation*}
\langle F(\varrho) \rangle_\varrho=\frac{3}{4\pi}\int_{\mathbb B} F[\rho(\vr)] dr_x dr_y dr_z\, .
\end{equation*}
We now define (for the outcome $\AI_+$) the average distance to be \footnote{Strictly speaking, we should average over states $\varrho$ with $\tr{\varrho\X(1)}\neq 0$. However, the complement set has  measure zero and does not therefore affect our calculations.}
\begin{equation*}
\dist^{\A,\X}_{+} := \left\langle\left[d^{\A,\X}_{+}(\varrho)\right]^2\right\rangle_{\varrho} \, .
\end{equation*}
We have chosen $\left[d^{\A,\X}_{+}(\varrho)\right]^2$ instead of $d^{\A,\X}_{+}(\varrho)$ just to simplify the calculations.

We can write \eqref{eq:d-bloch} as
\begin{eqnarray*}
\left[d^{\A,\X}_{+}(\varrho)\right]^2 & = & \frac{1}{4} \no{\vq_{++}+\vq_{+-}-2\vx}^2\\
& &+\frac{1}{4} f(\varrho)^2 \no{\vq_{++} - \vq_{+-}}^2 \\
& &+\frac{1}{2}f(\varrho) (\vq_{++}+\vq_{+-}-2\vx)\!\cdot\!(\vq_{++}-\vq_{+-}) .
\end{eqnarray*}
As $\langle f(\varrho)\rangle_\varrho=0$ and
\[
\langle f(\varrho)^2\rangle_\varrho=2-\frac{3\sqrt{2}}{2} \ln (1+\sqrt{2}) \equiv \alpha\, ,
\]
the average distance $\dist^{\A,\X}_{+}$ is expressible in the form
\begin{equation*}
\dist^{\A,\X}_{+}= \frac{1}{4} \no{\vq_{++}+\vq_{+-}-2\vx}^2 + \frac{\alpha}{4}  \no{\vq_{++} - \vq_{+-}}^2 \, .
\end{equation*}
Analogously, we get
\begin{eqnarray*}
\dist^{\B,\Y}_{+} & = &  \frac{1}{4} \no{\vq_{++} + \vq_{-+} - 2\vy}^2 + \frac{\alpha}{4}  \no{\vq_{++} - \vq_{-+}}^2 \, , \\
\dist^{\A,\X}_{-} & = &  \frac{1}{4} \no{\vq_{--} + \vq_{-+} + 2\vx}^2 + \frac{\alpha}{4}  \no{\vq_{--} - \vq_{-+}}^2 \, , \\
\dist^{\B,\Y}_{-} & = &  \frac{1}{4} \no{\vq_{--} + \vq_{+-} + 2\vy}^2 + \frac{\alpha}{4}  \no{\vq_{--} - \vq_{+-}}^2 \, .
\end{eqnarray*}

The distance $\dist^{\A,\X}_{+}$ can be made zero by taking $\vq_{\pm+}=\vq_{\pm-}=\pm\vx$.
This gives also $\dist^{\A,\X}_{-}=0$. But we will then have $\dist^{\B,\Y}_{\pm}=1+\alpha$ and the sum is then $\dist=\dist^{\A,\X}_{+}+\dist^{\A,\X}_{-}+\dist^{\B,\Y}_{+}+\dist^{\B,\Y}_{-}=2(1+\alpha)$. As we shall see this choice does not achieve
the minimal value of the sum $\dist$. 

We want to make all the four distances as small as possible, under the condition that they are equal.
To find the optimal instrument, we consider the sum 
$\dist= \dist^{\A,\X}_{+} + \dist^{\B,\Y}_{+} + \dist^{\A,\X}_{-} + \dist^{\B,\Y}_{-} ,
$ which is obviously a convex function of vectors $\vq_{\pm\pm}$.
Thus, its minimization subject to the conditions $\no{\vq_{\pm\pm}}\leq 1$ 
is a convex optimization problem. Since $\dist$ is differentiable, the optimality criterion (see e.g.~\cite{CO04}) for an instrument defined by a quadruple $\bar{\vq}=(\vq_{++},\vq_{+-},\vq_{-+},\vq_{--})$ is that
\begin{equation}\label{criterium}
\nabla \dist(\bar{\vq})^T\cdot \bar{\vp} \geq \nabla \dist(\bar{\vq})^T\cdot \bar{\vq}
\end{equation}
for all $\bar{\vp}=(\vp_{++},\vp_{+-},\vp_{-+},\vp_{--})$ with $\no{\vp_{\pm\pm}}\leq 1$.
It is now easy to verify that the optimal solution is achieved when $\vq_{\pm\pm}=(\pm\vx\pm\vy)/\sqrt{2}$ and this choice gives $ \dist^{\A,\X}_{+} = \dist^{\B,\Y}_{+} = \dist^{\A,\X}_{-} = \dist^{\B,\Y}_{-}=\frac{1}{2}(3-2\sqrt{2}+\alpha)$. 
We can also compare this solution with  the example where one direction was preferred, and we see that
\begin{equation*}
\dist_{\mathrm{opt}} = 2(3-2\sqrt{2}+\alpha) < 2(1+\alpha) \, .
\end{equation*}

The optimal joint instrument corresponds to the pure states 
$\xi_{\pm\pm}^{\rm ave}=2\G(\pm 1,\pm 1)$, 
and we recognize it being the L\"uders instrument 
$\Lu^\G$ of $\G$, given as
\begin{eqnarray}\nonumber
\Lu^\G_{\pm\pm}(\varrho) &=& \sqrt{\G(\pm1,\pm1)} \varrho \sqrt{\G(\pm1,\pm1)} \\
\label{eq:solution_1} &=& \tr{\G(\pm1,\pm1)\varrho}\xi_{\pm,\pm}^{\rm ave}\,.
\end{eqnarray}

In summary, we have found that the L\"uders instrument $\Lu^\G$ of $\G$ gives the optimal joint approximation of $\XI$ and $\YI$, the quality of approximations being quantified using the average distance. 

\subsection{Optimal approximations under the worst case distance}\label{sec:supoptimal}

The optimal joint instrument naturally depends on the quantification 
of the distance between two operations. We believe that the average norm 
studied in Sec.~\ref{sec:avgoptimal} is the most relevant way to 
measure the distance. 

If our task was to discriminate the given pair of instruments, the
average norm would then quantify the average success probability under
supposition of choosing the test state $\rho$ randomly. As a
comparison we take a look also on the \emph{worst-case distance},
which determines test states allowing the best possible
discrimination of the two instruments. In this sense the worst-case
distance optimizes the distinguishability over the test states. The
worst-case distance is defined as
\begin{equation*}
\sdist^{\A,\X}_{+} := \sup_{\varrho} d^{\A,\X}_{+}(\varrho) \, .
\end{equation*}
We want, again, to find an instrument which minimizes these distances under the condition that they are all equal.

Let us observe that the Bloch vector of a normalized outcome state $\AI_+(\varrho)/\tr{\AI_+(\varrho)}$ is of the form,
\begin{equation}\label{eq:output}
\frac{1}{2}[1+f(\varrho)]\vq_{++}+\frac{1}{2}[1-f(\varrho)]\vq_{+-}\, .
\end{equation}
Therefore, the distance  $d^{\A,\X}_{+}(\varrho)$ is the length of a vector being a convex combination of vectors $\vq_{++}-\vx$ and $\vq_{+-}-\vx$.
We thus conclude that
\begin{equation}\label{eq:dist_AP+}
\sdist^{\A,\X}_{+}=\max\left\{\no{\vq_{++}-\vx},\no{\vq_{+-}-\vx}\right\}\, .
\end{equation}
Similarly, we get
\begin{eqnarray*}
\sdist^{\B,\Y}_{+} & = & \max\left\{\no{\vq_{++}-\vy},\no{\vq_{-+}-\vy}\right\}, \label{eq:dist_BQ+}\\
\sdist^{\A,\X}_{-} & = & \max\left\{\no{\vq_{--}+\vx},\no{\vq_{-+}+\vx}\right\}, \label{eq:dist_AP-}\\
\sdist^{\B,\Y}_{-} & = & \max\left\{\no{\vq_{--}+\vy},\no{\vq_{+-}+\vy}\right\}. \label{eq:dist_BQ-}
\end{eqnarray*}

The distance $\sdist^{\A,\X}_{+}$ in \eqref{eq:dist_AP+} can be made zero, and this happens if and only if $\vq_{++}=\vq_{+-}=\vx$.
If $\vq_{++}$ and $\vq_{+-}$ are chosen in this way we are still free to choose $\vq_{-+}$ and $\vq_{--}$, hence we can also achieve $\sdist^{\A,\X}_{-}=0$ by taking $\vq_{--}=\vq_{-+}=-\vx$. However, we will then have $\sdist^{\B,\Y}_{\pm}=\sqrt{2}$ and $\sdist=\sdist^{\A,\X}_{+}+\sdist^{\B,\Y}_{+}+\sdist^{\A,\X}_{-}+\sdist^{\B,\Y}_{-}=2\sqrt{2}$. 

The value $\sdist=2\sqrt{2}$ for the sum is achieved also for the symmetric choice $\vq_{\pm\pm}=(\pm\vx\pm\vy)/2$. In this case all the distances are equal, $\sdist^{\A,\X}_{+}=\sdist^{\B,\Y}_{+}=\sdist^{\A,\X}_{-}=\sdist^{\B,\Y}_{-}=1/\sqrt{2}$.

Let us then minimize the distances under the condition that they are all equal. First, we require that $\sdist^{\A,\X}_{+}=\sdist^{\B,\Y}_{+}$ and we minimize these two distances, ignoring for a moment the other two distances. It is easy to see that in the optimal case it is necessary to choose $\vq_{++}=(\vx+\vy)/2$. Similarly, if we require that $\sdist^{\A,\X}_{-}=\sdist^{\B,\Y}_{-}$ and we minimize these two distances independently of the previous minimization, we see that it is necessary to put $\vq_{--}=(-\vx-\vy)/2$. These two optimal choices are possible simultaneously only if we set $\vq_{+-}=(\vx-\vy)/2$ and $\vq_{-+}=(-\vx+\vy)/2$. In conclusion, the symmetric choice $\vq_{\pm\pm}=(\pm\vx\pm\vy)/2$ is optimal.

The instrument $\GI$ corresponding to this symmetric choice of $\xi_{\pm\pm}$ can be written as
\begin{eqnarray}
\nonumber
\GI_{\pm\pm}(\varrho) &=& \frac{1}{\sqrt{2}} \Lu^\G_{\pm,\pm}(\varrho)+
\left(1-\frac{1}{\sqrt{2}}\right) 
\tr{\varrho\G(\pm1,\pm 1)} \half I\nonumber\\
\nonumber &=& \tr{\varrho \G(\pm1,\pm1)}\left[
\frac{1}{\sqrt{2}}\xi_{\pm\pm}^{\rm ave}+(1-\frac{1}{\sqrt{2}})\half I
\right]\\
&=& \label{eq:solution_2} \tr{\varrho \G(\pm1,\pm1)} \xi^{\rm w.c.}_{\pm\pm}\,.
\end{eqnarray}
Hence, $\GI$ is a mixture of the L\"uders instrument $\Lu^\G$ and another instrument, which has very simple form. 
In particular, $\GI$ can be realized by using the L\"uders instrument $\Lu^\G$, accepting the measurement outcomes but ignoring the output state in $1-\frac{1}{\sqrt{2}}$ parts of the measurement and preparing the maximally mixed state $\half\id$ in these cases. 

\begin{figure}
\begin{center}
\includegraphics[scale=0.85]{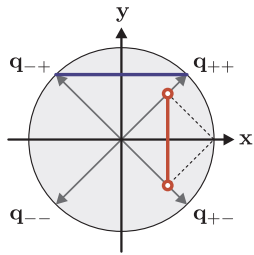}
\end{center}
\caption{\label{fig:2D} Bloch vector representation of instrument $\GI$ gives us in the case of approximation $\BI_+$ the resulting normalized output states lying on a (thick horizontal) blue line. The average distance is minimized for the choice of symmetrical pure states $\xi_{\pm\pm}$. In the case of approximation $\AI_+$ the normalized output states are represented by (thick vertical) red line with red dots representing the worst cases. These states are closest to $\vx$ when the output state is noisier than in the average case.}
\end{figure}

Previous results can be visualized as given in Fig.~\ref{fig:2D}. In the case of the average distance [horizontal blue line represents possible normalized output states as a manifestation of \eqref{eq:output}] we find it is natural to expect as large Bloch vectors $\vq_{\pm\pm}$ as possible, but for the worst case distance the additional noise in the normalized output state makes it closer to the desired state (vertical red line with red dots representing worst cases).

\section{Approximate joint measurement of three von Neumann measurements}\label{sec:three}

Let $\vz$ be a unit vector which is orthogonal to both $\vx$ and $\vy$, and let $\Z$ be the corresponding sharp observable.
Suppose we want to approximate the von Neumann instruments of $\X$, $\Y$, and $\Z$.
This problem of additional measurement bears some differences with the previously studied approximation task of two von Neumann measurements.

The optimal jointly measurable approximations of $\X$, $\Y$, and $\Z$ are given by
\begin{eqnarray*}\label{eq:ABC}
\A(\pm 1) &=& \frac{1}{2} (\id \pm \eta \sigma_\vx) \, ,\\
\B(\pm 1) &=& \frac{1}{2} (\id \pm \eta \sigma_\vy) \, ,\\
\C(\pm 1) &=& \frac{1}{2} (\id \pm \eta \sigma_\vz)  \,
\end{eqnarray*}
with $\eta=\frac{1}{\sqrt{3}}$.
The observables $\A$ and $\B$ are of similar form as before, but we have to decrease $\eta$ from  $\frac{1}{\sqrt{2}}$ to $\frac{1}{\sqrt{3}}$ to make it possible to include the additional spin component direction. It has been proved in \cite{BrAn07} that for $\eta>\frac{1}{\sqrt{3}}$ the three observables are not jointly measurable.

Generally, a joint observable $\E$ for $\A$, $\B$, and $\C$ has eight outcomes.
An observable defined as
\begin{equation*}
\E(\pm 1,\pm 1,\pm 1)=\frac{1}{8}\left[\id+\frac{1}{\sqrt{3}}(\pm\sigma_\vx\pm\sigma_\vy\pm\sigma_\vz)\right]
\end{equation*}
is a joint observable for $\A$, $\B$, and $\C$ since it satisfies the marginal conditions,
\begin{eqnarray*}
\A(\pm 1) &=& \sum_{j,k\in\{-1,1\}} \E(\pm 1,j,k)\, ,\\
\B(\pm 1) &=& \sum_{j,k\in\{-1,1\}} \E(j,\pm 1,k)\, ,\\
\C(\pm 1) &=& \sum_{j,k\in\{-1,1\}} \E(j,k,\pm 1)\, .
\end{eqnarray*}
Unlike in the earlier situation, now we have several different joint observables. 
Another joint observable $\F$ for $\A$, $\B$, and $\C$ is given by
\begin{eqnarray*}
\F(1,1,1) &=&2\E(1,1,1),\\
\F(1,-1,-1) &=& 2\E(1,-1,-1),\\
\F(-1,-1,1) &=& 2\E(-1,-1,1),\\
\F(-1,1,-1) &=& 2\E(-1,1,-1),\\
\F(1,1,-1) &=& \F(1,-1,1)=\F(-1,1,1)\\
&=& \F(-1,-1,-1)=0 \, .
\end{eqnarray*}

A notable feature of the observable $\F$ is that it is essentially a four-outcome
observable. Although $\A$, $\B$, and $\C$ have infinitely many different joint observables, $\F$ is the unique joint observable having only four nonzero 
elements. To demonstrate this fact, suppose that $\F'$ is a joint observable of $\A$, $\B$, and $\C$, and that
\begin{eqnarray*}
 \F'(1,1,-1) &=& \F'(1,-1,1)=\F'(-1,1,1)\\
 &=& \F'(-1,-1,-1)=0 \, .
\end{eqnarray*}
First of all, let us notice that $\F'$ is completely determined by a single nonzero element, say $\F'(1,1,1)$. 
The other operators are given by the marginal conditions.
For instance, $\F'(1,-1-1)=\A(1)-\F'(1,1,1)$.
Since $\F'$ is an observable, the operators $\F'(i,j,k)$ must sum up to identity.
Hence, we get
\begin{eqnarray*}
\id &=& \F'(1,1,1)+\F'(1,-1,-1)+\F'(-1,1,-1)\\
&& +\F'(-1,-1,1)\\
&=& \F'(1,1,1)+[\A(1)-\F'(1,1,1)]+[\B(1)-\F'(1,1,1)]\\
&& +[\C(1)-\F'(1,1,1)] \, .
\end{eqnarray*}
Therefore, the operator $\F'(1,1,1)$ is determined by the equation,
\begin{equation*}
\F'(1,1,1)=\frac{1}{2}[\A(1)+\B(1)+\C(1)-\id] \, .
\end{equation*}
It follows that $\F'=\F$, and the four-outcome joint observable is hence unique.

As done previously for the joint observable $\G$, we can now study the instruments implementing $\E$ and $\F$.
The elements forming $\E$ and $\F$ are rank-1 operators.
Therefore, our characterization for instruments implementing $\G$ in Sec.~\ref{sec:joint-instrument} applies to these two observables as well.

An instrument implementing $\E$ is determined by eight
states $\zeta_{\pm,\pm,\pm}$ and the corresponding operations are
\begin{equation*}
\EI_{\pm\pm\pm}(\varrho)=\tr {\varrho \E(\pm1,\pm1,\pm1)}\zeta_{\pm\pm\pm}
\end{equation*}
The approximations to von Neumann instruments $\XI$, $\YI$, and $\ZI$ are now defined as follows:
\begin{eqnarray*}
\label{eq:approx_P8+}\AI_{\pm} \equiv \EI_{\pm++} + \EI_{\pm-+}+ \EI_{\pm--}+ \EI_{\pm+-} & \!\sim\! & \XI_{\pm},\\
\label{eq:approx_Q8+}\BI_{\pm} \equiv \EI_{+\pm+} + \EI_{-\pm+}+ \EI_{+\pm-}+ \EI_{-\pm-} & \!\sim\! & \YI_{\pm},\\
\label{eq:approx_R8+}\CI_{\pm} \equiv \EI_{++\pm} + \EI_{--\pm}+ \EI_{+-\pm}+ \EI_{-+\pm} & \!\sim\! & \ZI_{\pm}.
\end{eqnarray*}
The optimal instrument under the average distance can be deduced by following a similar procedure as the one presented in Sec.~\ref{sec:avgoptimal}.
We thus determine the minimum of the sum $\dist=\dist^{\A,\X}_{+}+\dist^{\B,\Y}_{+}+\dist^{\C,\Z}_{+}+\dist^{\A,\X}_{-}+\dist^{\B,\Y}_{-}+\dist^{\C,\Z}_{-}$   where now,  for instance, the distance $\dist^{\A,\X}_{+}$ is
\begin{eqnarray*}
\dist^{\A,\X}_{+}&=& \frac{1}{16} 
\no{\vq_{+++}+\vq_{+-+}+\vq_{+--}+\vq_{++-}-4\vx}^2 \nonumber
\\\nonumber & & 
+\frac{\beta}{16}\no{\vq_{+++} -\vq_{+-+} -\vq_{+--} +\vq_{++-}}^2
\\ & &
+\frac{\beta}{16}\no{\vq_{+++} + \vq_{+-+} - \vq_{+--} - \vq_{++-}}^2
 \end{eqnarray*}
and
\[
\beta \equiv \left\langle \left(\frac{r_y}{\sqrt{3}+r_x}\right)^2 \right\rangle_\varrho = \half [ 7-3\sqrt{3} \ln (2+\sqrt{3})]\, .
\]
Using the criterion \eqref{criterium}, it is straightforward to verify that the optimal solution is achieved when $\vq_{\pm\pm\pm}=\frac{1}{\sqrt{3}}\left(\pm \vx\pm\vy\pm\vz\right)$. 
All the distances are then equal and take the value $\frac{2}{3}\beta+(1-\frac{1}{\sqrt{3}})^2$.
The related joint instrument is the L\"uders instrument $\Lu^\E$ of $\E$. 

An instrument $\FI$ implementing $\F$ is, again, determined by eight states. 
However, the states corresponding to the zero elements play no role, so $\FI$ is actually determined by four states only.
Since $\F$ gives different measurement outcome probabilities than $\E$, the related average distances are also different. 
So, if we consider $\F$, we get
\begin{equation*}
\dist^{\A,\X}_{+}=\frac{1}{4} \no{\vq_{+++}+\vq_{+--}-2\vx}^2 + \frac{\gamma}{4}  \no{\vq_{+++} - \vq_{+--}}^2,
\end{equation*}
and
\[
\gamma \equiv \left\langle \left(\frac{r_y+r_z}{\sqrt{3}+r_x}\right)^2 \right\rangle_\varrho = 2\beta \, .
\]
The sum $\dist=\dist^{\A,\X}_{+} + \dist^{\B,\Y}_{+} + \dist^{\C,\Z}_{+} +\dist^{\A,\X}_{-} + \dist^{\B,\Y}_{-}+  \dist^{\C,\Z}_{-} $
achieves its minimum when the four relevant Bloch vectors are
\begin{eqnarray*}
\vq_{+++} &=& (\vx+\vy+\vz)/\sqrt{3},\\
\vq_{+--} &=& (\vx-\vy-\vz)/\sqrt{3},\\
\vq_{-+-} &=& (-\vx+\vy-\vz)/\sqrt{3},\\
\vq_{--+} &=& (-\vx-\vy+\vz)/\sqrt{3}.
\end{eqnarray*}
This choice gives $\dist^{\A,\X}_{+}=\frac{2}{3}\gamma+\left(1-\frac{1}{\sqrt{3}}\right)^2$.
The related joint instrument is the L\"uders instrument $\Lu^\F$ of $\F$. 

\begin{figure}
\begin{center}
\includegraphics[scale=1]{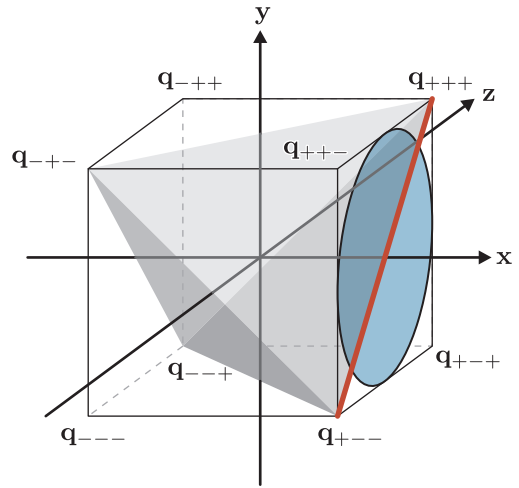}
\end{center}
\caption{\label{fig:3D} Bloch vector comparison of the four- and eight-outcome instruments. The four-outcome instrument leads to normalized output states from the edges of the tetrahedron (especially for the measurement along the $x$ axis illustrated by the thick red diagonal). The eight-outcome instrument leads to the faces of the cube given by the vectors $\vq_{\pm\pm\pm}$ (especially for the measurement along the $x$ axis illustrated by the blue circle).}
\end{figure}


We can, again, use Bloch representation and write the normalized output states similarly as in \eqref{eq:output}. For the four-outcome instrument we get a similar result --- the normalized output states are convex combinations of the corresponding states defining the instrument. For instance, the normalized output state, when measuring along axis $x$ and obtaining the outcome $+1$, is a convex combination of $\vq_{+++}$ and $\vq_{+--}$. Hence, these Bloch vectors of output states  lie on the edges of a tetrahedron as depicted on Fig.~\ref{fig:3D} (thick red line).

For the eight-outcome instrument we find that the corresponding output states form a set
\begin{equation*}
\frac{1}{\sqrt{3}}(\vx+\tilde r_y\vy+\tilde r_z\vz),\quad \tilde{r}_y^2+ \tilde{r}_z^2\leq 1.
\end{equation*}
This set is a circle lying in the plane given by the vectors $\vq_{+jk}$, $j,k\in\{-1,1\}$ [i.e.,~being inscribed into the face of the cube such as is depicted in Fig.~\ref{fig:3D} (blue circle)]. From this geometrical representation we can confirm the fact shown earlier;  under the usage of the average distance the four-outcome instrument is worse that the eight-outcome instrument as it has a contribution from states being further from the reference state.

\section{Conclusions}

The impossibility of joint measurements of orthogonal qubit measurements leads us to the study of their approximations. 
We have considered not only observables, but we have accessed the problem from the perspective of instruments.
We have characterized the optimal approximation of two von Neumann instruments, and the optimal joint instrument was found 
to be the L\"{u}ders instrument of the optimal (unique) joint observable. 
This result [see Eq.~\eqref{eq:solution_1}] was achieved by searching for the best approximation under 
the average distance between the normalized output states. 
When considering the worst case norm, the resulting optimal instrument is a mixture of the corresponding L\"{u}ders instrument 
and the state-space contraction into the complete mixture [see Eq.~\eqref{eq:solution_2}].

A similar investigation was performed in the case of three von Neumann instruments, but then we faced the problem of non-uniqueness 
of the optimal joint observable. Nevertheless, in the two commonly used instances the L\"{u}ders instrument was found to be optimal.
Although the numerical value of the minimal average distance does not have any intrinsic meaning, one  can use it in order to compare the minimal distances in the three studied cases: approximation of two von Neumann instruments and approximations of three von Neumann 
instruments with eight- and four-outcome measurements. The minimal average distances for these approximations are $0.15$, $0.23$, and $0.28$, respectively.
As one would expect, the average distance can be made lowest in the first case.
Namely, it is certainly easier to approximate two von Neumann instruments rather than three.
Furthermore, it is not surprising that in the latter two cases eight instead of four outcomes are more efficient in the approximation task. This may be explained by a broader set of outcomes to choose from, leaving less space for error.

We believe that L\"{u}ders instruments are optimal joint instruments for a more general class of situations than only those studied here. 
A natural extension of the two orthogonal qubit observables would be the case of two sharp observables related to mutually unbiased bases.
Another interesting class is that of the continuous variable systems, where phase space observables play the role of joint observables.
These generalizations merit further study.

\section*{Acknowledgments}
T.H. and M.A.J. acknowledge financial support from the Danish National Research Foundation Center for Quantum Optics (QUANTOP) and from the European Union projects COQUIT and QUEVADIS.
D.R. and M.Z. acknowledge financial support from the European Union Project
No.~HIP FP7-ICT-2007-C-221889, and from Projects
No.~APVV-0673-07 QIAM, No.~OP CE QUTE ITMS NFP 262401022,
and No.~CE-SAS QUTE. M.Z.~also acknowledges support from Project No.~MSM0021622419.
The authors thank Peter Stano for useful comments.


\begin{thebibliography}{10}

\bibitem{Busch86}
P.~Busch.
\newblock {\em Phys. Rev. D}, {\bf 33,} 2253 (1986).

\bibitem{BuSh06}
P.~Busch and C.~Shilladay.
\newblock {\em Phys. Rep.} {\bf 435,} 1 (2006).

\bibitem{LiLiYuCh09}
N.-L.~Liu, L.~Li, S.~Yu, and Z.-B.~Chen.
\newblock {\em Phys. Rev. A} {\bf 79} 052108 (2009).

\bibitem{KuSaUe07}
Y.~Kurotani, T.~Sagawa, and M.~Ueda.
\newblock {\em Phys. Rev. A} {\bf 76} 022325 (2007).

\bibitem{SaUe08}
T.~Sagawa and M.~Ueda.
\newblock {\em Phys. Rev. A} {\bf 77,} 012313 (2008).

\bibitem{BuHe08}
P.~Busch and T.~Heinosaari.
\newblock {\em Quant. Inf. Comp.} {\bf 8,} 0797 (2008).

\bibitem{BrAnBa09}
T.~Brougham, E.~Andersson, and S.M.~Barnett.
\newblock {\em Phys. Rev. A} {\bf 80,} 042106 (2009).

\bibitem{StReHe08}
P.~Stano, D.~Reitzner, and T.~Heinosaari.
\newblock {\em Phys. Rev. A} {\bf 78,} 012315 (2008).

\bibitem{YuLiLiOh08}
S.~Yu, N.~Liu, L.~Li, and C.H.~Oh.
\newblock arXiv:0805.1538v1 [quant-ph], 2008.

\bibitem{BuSc10}
P.~Busch and H.-J.~Schmidt.
\newblock {\em Quantum Inf. Process.} {\bf 9,} 143 (2010).

\bibitem{ClHoShHo69}
J.F.~Clauser, M.A.~Horne, A.~Shimony, and R.A.~Holt.
\newblock {\em Phys. Rev. Lett.} {\bf 23,} 880 (1969).

\bibitem{AnBaAs05}
E.~Andersson, S.M.~Barnett, and A.~Aspect.
\newblock {\em Phys. Rev. A} {\bf 72,} 042104 (2005).

\bibitem{WoPeFe09}
M.M.~Wolf, D.~Perez-Garcia, and C.~Fernandez.
\newblock {\em Phys. Rev. Lett.} {\bf 103,} 230402 (2009).

\bibitem{FePa07}
A.~Ferraro and M.G.A.~Paris.
\newblock {\em Open Sys. \& Information Dyn.} {\bf 14,} 149 (2007).


\bibitem{BuLa95}
P.~Busch and P.~Lahti.
\newblock {\em Riv. Nuovo Cimento} {\bf 18,} 1 (1995), e-print arXiv:quant-ph/0406132v1.

\bibitem{PSAQT82}
A.S.~Holevo.
\newblock {\em Probabilistic and Statistical Aspects of Quantum Theory}.
\newblock (North-Holland Publishing Co., Amsterdam, 1982).

\bibitem{OQP97}
P.~Busch, M.~Grabowski, and P.J.~Lahti.
\newblock {\em Operational Quantum Physics}.
\newblock (Springer-Verlag, Berlin, 1997),
\newblock 2nd corrected printing.

\bibitem{Busch87}
P.~Busch.
\newblock {\em Found. Phys.} {\bf 17,} 905 (1987).

\bibitem{QTOS76}
E.B.~Davies.
\newblock {\em Quantum Theory of Open Systems}.
\newblock (Academic Press, London, 1976).

\bibitem{Com}
P.~Busch and P.~Lahti.
\newblock {\em L\"uders Rule}.
\newblock in {\em Compendium of Quantum Physics,} editored by D.~Greenberger, K.~Hentschel, and F.~Weinert (Springer, New York, 2009).


\bibitem{HeReStZi09}
T.~Heinosaari, D.~Reitzner, P.~Stano, and M.~Ziman.
\newblock {\em J. Phys. A} {\bf 42,} 365302 (2009).

\bibitem{QDET76}
C.W.~Helstrom.
\newblock {\em Quantum Detection and Estimation Theory}
\newblock (Academic Press, New York, 1976).

\bibitem{CO04}
S.~Boyd and L.~Vandenberghe.
\newblock {\em Convex Optimization}
\newblock (Cambridge University Press, Cambridge, 2004).

\bibitem{BrAn07}
T.~Brougham and E.~Andersson.
\newblock {\em Phys. Rev. A} {\bf 76,} 052313 (2007).

\end{thebibliography}
\end{document}